\documentclass[pre,twocolumn,superscriptaddress]{revtex4}
\usepackage{epsfig,amsmath,revsymb}
\begin {document}
\title {Exponential velocity tails in a driven inelastic Maxwell model}
\author{T. Antal} \email{Tibor.Antal@physics.unige.ch}
\affiliation{Department of Physics, University of Geneva, CH 1211
Geneva 4, Switzerland}  
\affiliation{Institute for Theoretical Physics, E\"otv\"os University,
1117 Budapest, P\'azm\'any s\'et\'any 1/a, Hungary}
\author{Michel Droz}\email{Michel.Droz@physics.unige.ch}
\affiliation{Department of Physics, University of Geneva, CH 1211
Geneva 4, Switzerland}  
\author{Adam Lipowski}\email{lipowski@amu.edu.pl}
\affiliation{Department of Physics, University of Geneva, CH 1211
Geneva 4, Switzerland}
\affiliation{Department of Physics, Adam Mickiewicz University,
61-614 Pozna\'{n}, Poland}
\begin {abstract}
The problem of the steady-state velocity distribution in a driven inelastic Maxwell
model of shaken granular material is revisited.
Numerical solution of the master equation and analytical arguments 
show that the model has 
bilateral exponential velocity tails ($P(v)\sim e^{-|v|/\sqrt D}$), where $D$ is 
the amplitude of the noise.
Previous study of this model predicted Gaussian tails  
($P(v)\sim e^{-av^2}$).
\end{abstract}
\maketitle
Recently, granular systems have been intensively studied~\mbox{\cite{SWINNEY,MUZZIO}}.
One reason of such an interest is the fact that in many respects these
systems are highly unconventional and even exotic~\cite{KADANOFF}.
As a result, very often even standard laws of statistical physics require certain
modification when applied to granular systems.

As an example, let us consider the Maxwell-Boltzmann law for velocity 
distribution $P(v)$ of atoms or molecules in a gaseous state, which states that 
$P(v)\sim {\rm e}^{-av^2}$.
Under certain experimental conditions, a granular system can be considered as a
gas and a natural question is what is its velocity distribution.
Numerous theoretical and experimental studies do not provide a simple answer to 
this question.
On the contrary, they show that $P(v)$ depends on certain details of the 
system as for example how energy is transferred by a thermostat into the system, 
in order to balance the energy loss during inelastic collisions.
Theoretical works have shown that $P(v)$ might be of the form 
${\rm e}^{-av^2}$~\cite{SWINNEY1,BIBEN},
${\rm e}^{-av^{3/2}}$~\cite{NOIJE,BARRAT} or 
${\rm e}^{-av}$~\cite{MONTANERO}.
Under certain conditions experiments show clear deviations from the
Maxwell-Boltzmann law, but it is still rather difficult to decide what is the form
of $P(v)$ in real granular systems~\cite{OLAFSEN}.

Very often granular systems are described using the so-called inelastic 
hard-sphere models, which could be then analysed using corresponding Boltzmann 
equations.
These off-lattice, two- or three-dimensional systems are however very difficult to 
study, especially when we want to explore large-velocity regions of the phase space.
A possible alternative is to construct simplified models for which more 
accurate calculations are possible.
One class of such models are Maxwell models, for which the collision term 
in the corresponding Boltzmann equation is velocity independent.
Recent studies show that for Maxwell models $P(v)$ might take either 
exponential (${\rm e}^{-av}$)~\cite{ERNST} or Gaussian 
(${\rm e}^{-av^2}$) form~\cite{BEN,CARILLO}.

A particularly interesting Maxwell model was proposed by Ben-Naim and 
Krapivsky (BN-K)~\cite{BEN}.
These authors presented an elegant solution of the master equation of their model using a
Fourier transform method.
However, their conclusion that $P(v)$ has a Gaussian decay is based on certain
approximation whose validity is difficult to assess.
Namely, they infer the large-velocity behaviour of $P(v)$ from the low-$k$ behaviour of 
the Fourier transform $\hat P(k)$.
Since BN-K model is one of the very few models for which exact or numerical but 
very precise calculations can be made, it would be desirable to clarify the validity
of this approach.

In the present paper we reexamine the BN-K model.
Analysing numerically the solution of the master equation of the model we obtain the velocity
distribution $P(v)$.
Asymptotically ($v\rightarrow\infty$), this quantity shows bilateral exponential tails 
($P(v)\sim {\rm e}^{-a|v|}$) and such a behaviour is seen over more than 10 decades.
We also present analytical arguments which support existence of 
bilateral exponential tails in this model.
From our analysis it follows that $a=\frac{1}{\sqrt D}$, where $D$ is the 
amplitude of noise which simulates the input of energy into the system.

To introduce the model, we consider a collection of particles that are 
characterized by a single parameter, their velocity $v$.
In this model positions of particles are not specified hence the model neglects any
spatial correlations.
Particles undergo two-body inelastic collisions that change their velocities 
according to $(v_1,v_2)\rightarrow(v_1',v_2')$, where
\begin{equation}
\begin{pmatrix} v_1' \\ v_2'\end{pmatrix} = 
\begin{pmatrix} \gamma & 1-\gamma \\ 1-\gamma & \gamma \end{pmatrix}
\begin{pmatrix} v_1 \\ v_2\end{pmatrix},
\label{e1}
\end{equation}
and  $\gamma$ is the inelasticity parameter ($0<\gamma<1$).
Particles which participate in a collision are chosen at random.
In addition to the collision rule (\ref{e1}), particles are subjected to the
uncorrelated white noise of strength $D$.
Existence of noise ensures that the model has a well-defined nontrivial steady 
state.
In the steady-state the velocity distribution $P(v)$ satisfies
the following equation~\cite{BEN}
\begin{equation}
-DP''(v)=-P(v)+
\frac{1}{1-\gamma}\int_{-\infty}^{+\infty} duP(u)
P\left(\frac{v-\gamma u}{1-\gamma}\right),
\label{e2}
\end{equation}
where double prime $''$ denotes the second derivative with respect to velocity.
To solve eq.~(\ref{e2}) BN-K introduced the Fourier transform of the velocity 
distribution $\hat P(k) = \int dv e^{ikv} P(v)$, which
in the steady state satisfies
\begin{equation}
(1+Dk^2)\hat P(k) = \hat P[\gamma k] \hat P[(1-\gamma)k].
\label{pksteady}
\end{equation}
This equation admits the following solution
\begin{equation}
\hat P(k) = \prod_{i=0}^{\infty} \prod_{j=0}^i [1+\gamma^{2j}(1-\gamma)^{2(i-j)}Dk^2].
\label{pkrecur}
\end{equation}
After some transformations eq.~(\ref{pkrecur}) can be written as
\begin{equation}
\hat P(k) = {\rm exp} \left[ \sum_{n=1}^\infty 
\frac{(-Dk^2)^n}{na_{2n}(\gamma)} \right] ~,
\label{final}
\end{equation}
where $a_n(\gamma) = 1 - (1-\gamma)^n - \gamma^n$.
To obtain  a large velocity behaviour of $P(v)$ BN-K truncate the series in 
eq.~(\ref{final}) keeping only the first, quadratic in $k$, term
\begin{equation}
\hat P(k)\approx {\rm exp}\left[\frac{-Dk^2}{a_2(\gamma)}\right].
\label{app}
\end{equation}
Subsequently, taking the inverse Fourier transform of this Gaussian function they obtain 
$P(v)\sim {\rm exp}(\frac{-v^2}{2v_0^2})$, where $v_0^2=2D/a_2(\gamma)$.

In our opinion, the above procedure of truncating the series and inverting the resulting 
function is not well justified.
The approximation (\ref{app}) agrees with the exact 
expression (\ref{final}) only up to the $k^2$ term of the Taylor expansion.
Equally consistent approximation can be given as
\begin{equation}
\hat P(k)\approx \frac{c^2}{c^2+k^2},
\label{app1}
\end{equation}
where $c=\sqrt{\frac{a_2(\gamma)}{D}}$.
But the inverse Fourier transform of~(\ref{app1}) is $\frac{c}{2}{\rm exp}(-c|v|)$, which
is qualitatively different than the Gaussian decay obtained by BN-K.
Both (\ref{app}) and (\ref{app1}) are consistent with the exact solution (\ref{final}) up
to the $k^2$ order  but differ at the higher orders.
Apparently, these higher order terms qualitatively affect the large-velocity limit of 
$P(v)$.
Consequently, without more detailed arguments such approximations are not justified.
Indeed, as we will show below, $P(v)$ has a bilateral exponential decay but with a 
different coefficient $c$.

\begin{figure}
\centerline{\epsfxsize=9cm 
\epsfbox{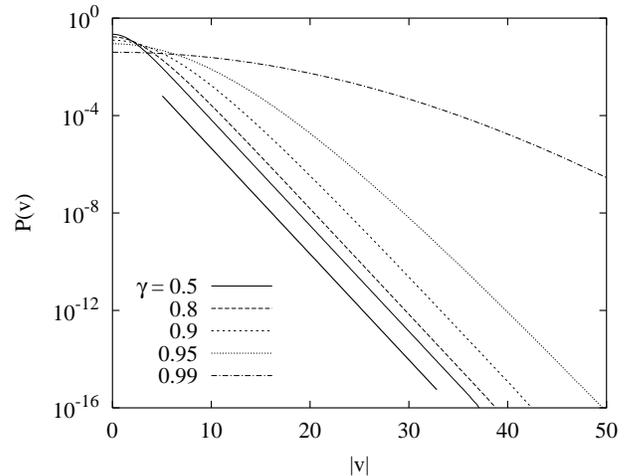}
}
\caption{
The velocity distribution $P(v)$ as a function of $v$ calculated for $D=1$.
Due to symmetry only $v>0$ part is shown.
The thick line represents the exponential function $\exp (-v)$.
}
\label{fft}
\end{figure}

To check the validity of the BN-K approach we calculate $P(v)$ numerically as an inverse
Fourier transform of $\hat P(k)$.
One way to compute $\hat P(k)$ is to evaluate the infinite sums in the logarithm of
eq.~(\ref{pkrecur}).
However, it turns out that using eq.~(\ref{final}) leads to a better precision.
Let us notice, however, that the series in eq.~(\ref{final}) converges but only for
$|k|<k_c=\frac{1}{\sqrt D}$.
To calculate $\hat P(k)$ for $|k|>k_c$ we can then use eq.~(\ref{pksteady}), 
provided that $\gamma k$ and $(1-\gamma)k$ fall within the range of convergence.
If not, we have to calculate $\hat P(\gamma k)$ and $\hat P((1-\gamma)k)$
referring again to eq.~(\ref{pksteady}).
Implementing this recursive procedure, we calculated $\hat P(k)$ and then using the 
Fast Fourier Transformation algorithm we obtained $P(v)$.
Our results for $D=1$ are shown in Fig.~\ref{fft}.
This figure clearly shows that $P(v)$ decays exponentially.
For $\gamma=0.5$ and 0.8 such a behaviour is seen for more than 10 decades.
For $\gamma$ close to unity there is a $\exp (-av^2)$ decay for small velocities which 
asymptotically is replaced by the bilateral exponential decay ($\exp (-c|v|)$).
Moreover, as shown in Fig.~\ref{fft} the asymptotic slope $c$ is independent of 
$\gamma$ and is approximately one.
As we will show below this slope only depends on the amplitude of noise $D$.
Before doing that let us notice that the $D$-dependence of the velocity distribution
$P(v)$ can be easily inferred from the fact that $D$ and $k$ enter its Fourier transform
only through $Dk^2$ terms (see eqs.~(\ref{pkrecur}) and (\ref{final})).
From this property one can easily obtain that
\begin{equation}
P(v, D) = \frac{1}{\sqrt{D}} P\left(\frac{v}{\sqrt{D}},D=1 \right),
\label{rescal}
\end{equation}
where we explicitly indicated the dependence on the noise $D$.

In the following we provide some analytical arguments showing 
that $P(v)$ decays exponentially.
Let us assume that in eq.~(\ref{e2}) the second term (gain) can be neglected.
Then eq.~(\ref{e2}) simplifies to $DP''(v)=P(v)$ and the normalizable solution reads
\begin{equation}
P(v)\sim e^{-\frac{|v|}{\sqrt D}}.
\label{e3}
\end{equation}
Let us notice that the exponential part is $\gamma$-independent 
(since $\gamma$ enters the master equation only through the neglected 
gain term).
Moreover, such a solution is in a very good agreement with numerical 
calculations (see Fig.~\ref{fft}).

For more general models it is known that neglecting the gain term is justified 
for $v\rightarrow\infty$ when the resulting solution decays faster than 
exponentially~\cite{NOIJE,BARRAT}.
Our solution (\ref{e3}) is thus a marginal case.
However, we can show that for our model the gain term in the limit 
$v\rightarrow\infty$ indeed can be neglected.
First, let us evaluate the gain term for the solution (\ref{e3}).
Elementary integration for $D=1$ and $v>0$ gives
\begin{multline}
\frac{1}{1-\gamma}\int_{-\infty}^{+\infty} du
{\rm exp}(-|u|)
{\rm exp}\Bigl[-\Big|\frac{v-\gamma u}{1-\gamma}\Big|\Bigr] \\ =
\frac{2}{2\gamma-1}\left[\gamma {\rm exp} \left(\frac{-v}{\gamma}\right)
+(\gamma-1){\rm exp} \left(\frac{-v}{1-\gamma}\right)\right]
\label{e4}
\end{multline}
One can see that since $0<\gamma<1$, the gain term decays exponentially 
with $v$ but faster than the solution (\ref{e3}) (i.e., a non-neglected loss term 
in (\ref{e2})).
We expect that eq.~(\ref{e3}) is only an asymptotic ($v\rightarrow\infty$) 
solution of the master equation (\ref{e2}).
Thus, for velocities $u\sim O(1)$ the distribution $P(u)$  deviates
from the asymptotic form (\ref{e3}).
This will modify the integral (\ref{e4}) but only in the vicinity of 
$u=0$ and $u=\frac{v}{\gamma}$.
As we argue below, such a modification of $P(u)$ still leads to the gain term decaying
faster than the loss term.

Indeed, the contribution around $u=0$ is a product of $P(u)$ and of the 
exponential term ${\rm exp}\bigl(\frac{-v-\gamma u}{1-\gamma}\bigr)$.
Thus, for large $v$ a modification of $P(u)$ around $u=0$ will change our
estimation (\ref{e4}) but only at the order of 
${\rm exp}(\frac{-v}{1-\gamma})$.
Similarly, one can show that a modification of $P(u)$ around 
$u=\frac{v}{\gamma}$ will change (\ref{e4}) by a factor of the order of 
${\rm exp}(\frac{-v}{\gamma})$.
Consequently, the gain term again decays faster than the solution $P(v)$,
which justifies its neglect.
Let us also notice that when $\gamma$ approaches 0 or 1 the model becomes
energy conserving and the bilateral exponential distribution (\ref{e3}) is no longer 
expected to hold.
But it is easy to notice that in such cases the decay of the gain term  
matches the decay of the solution $P(v)$ and it cannot be neglected.

As a comment let us notice that knowing the second moment of $P(v)$~\cite{BEN} 
enables us to calculate the effective temperature $T$ of our systems 
defined as an averaged square velocity
\begin{equation}
T=\langle v^2\rangle = \frac{D}{\gamma(1-\gamma)}.
\label{temp}
\end{equation}
\begin{figure}
\centerline{\epsfxsize=9cm 
\epsfbox{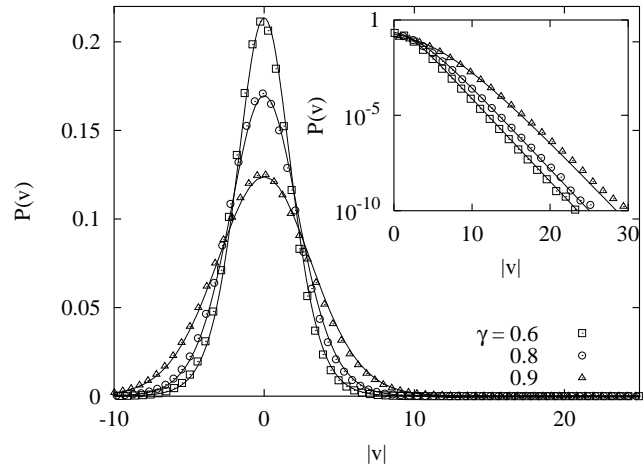}
}
\caption{
Velocity distribution $P(v)$ calculated using  Monte Carlo simulations.
Simulations were made for $N=10^5$ particles.
Continuous lines are the results obtained from Fourier inversion.
The inset shows our data in the semi-logarithmic scale.
}
\label{monte}
\end{figure} 
To check the validity of our calculations we also made Monte Carlo simulations of this
model.
Results, which are shown in Fig.~\ref{monte} confirm the bilateral exponential 
decay of $P(v)$ 
although the accuracy is this time much lower.
Our Monte Carlo data are rescaled in such a way that their variance matches that 
obtained using the Fourier transform method and a very good agreement is seen even on
the logarithmic scale.

In conclusion, we have shown that the Maxwell model proposed by Ben-Naim and Krapivsky 
has velocity distribution decaying as a bilateral exponential.
Together with the recent results by Ernst and Brito~\cite{ERNST} it indicates that 
such a decay might 
be of more generic nature for this class of models.
As a possible extension, it would be desirable to examine some other Maxwell models
where velocities are not scalars but rather $d$-dimensional vectors.
Actually, such models were already studied and the analysis indicates that, for 
increasing
$d$, correlations between velocities and deviation from the pure Gaussian distribution 
decrease~\cite{BEN1}.
One possibility is that there might be a critical dimension $d_c$ and such that 
for $d<d_c$ the velocity distribution has a bilateral exponential decay (as in the 
present model) while it it has  Gaussian
decay for $d\geq d_c$.
Analysis of such models is however left for the future.
\begin{acknowledgments} 
We acknowledge interesting discussion with Fran\c{c}ois Coppex.
This work was partially supported by the Swiss National Science Foundation
and the project OFES 00-0578 "COSYC OF SENS" and 
the Hungarian Academy of Sciences (Grant No. OTKA T029792).
\end{acknowledgments}

\end {document}